\documentclass[12pt]{iopart}

\usepackage{graphicx}
\usepackage{subfigure}

\begin{document}

\title[]{Particle dynamics of a cartoon dune}

\author{Christopher Groh$^1$, Ingo Rehberg$^1$, and Christof A. Kruelle$^{1,2}$}
\address{$^1$Experimentalphysik V, Universit\"{a}t Bayreuth, D-95440 Bayreuth, Germany}
\address{$^2$Maschinenbau und Mechatronik, Hochschule Karlsruhe - Technik und Wirtschaft, D-76133 Karlsruhe, Germany}

\begin{abstract}
The spatio-temporal evolution of a downsized model for a desert dune is observed experimentally in a narrow water flow channel. A particle tracking method reveals that the migration speed of the model dune is one order of magnitude smaller than that of individual grains. In particular, the erosion rate consists of comparable contributions from creeping (low energy) and saltating (high energy) particles. The saltation flow rate is slightly larger, whereas the number of saltating particles is one order of magnitude lower than that of the creeping ones. The velocity field of the saltating particles is comparable to the velocity field of the driving fluid. It can be observed that the spatial profile of the shear stress reaches its maximum value upstream of the crest, while its minimum lies at the downstream foot of the dune. The particle tracking method reveals that the deposition of entrained particles occurs primarily in the region between these two extrema of the shear stress. Moreover, it is demonstrated that the initial triangular heap evolves to a steady state with constant mass, shape, velocity, and packing fraction after one turnover time has elapsed. Within that time the mean distance between particles initially in contact reaches a value of approximately one quarter of the dune basis length.
\end{abstract}

\maketitle

`` `We are right in the open desert,' said the doctor. `Look at that vast reach of sand! What a strange spectacle! What a singular arrangement of nature!' " \cite{verne1863}. In this fictive story of Jules Verne from the 19th century, Dr.\,Samuel Ferguson and his two companions discover the beauty of sand dunes in the deserts of Africa by traveling five weeks in a balloon across the continent. Today, scientist are still overwhelmed by these self-organized granular structures, but use satellites for observations instead of balloons \cite{andreotti2009,nasa2009}.

The more descriptive science of the past century \cite{bagnold1941, pye1990, lancaster1995} has changed to a detailed approach in understanding the dune dynamics \cite{andreotti2002a}. The crescent-shaped \textit{barchan} dune was chosen as a suitable object because of its relatively fast dynamics. So-called `minimal models' were established to describe the basic dynamics of barchan dunes \cite{andreotti2002b, kroy2002a, kroy2002b}. These two-dimensional models deal with dune slices along the direction of the driving wind. Shortly speaking, they combine an analytical description of the turbulent shear flow over low elevations \cite{hunt1988, weng1991} with a continuum description, which models the saltation on the surface of the dune \cite{sauermann2001}. In the next step of complexity, these two-dimensional slices are coupled in the cross-wind direction to model three-dimensional barchan dunes \cite{schwaemmle2003}. Laboratory experiments \cite{andreotti2002a, endo2004, endo2005, hersen2002, hersen2005} and field measurements on Earth \cite{sauermann2000, elbelrhiti2005, elbelrhiti2008} or satellite observations of Mars \cite{parteli2007a, parteli2007b} reveal the quality of such models. Some aspects of the minimal models have been checked quantitatively in a narrow water flow channel \cite{groh2008, groh2009b, groh2009c}, and the existence of a shape attractor for barchan dunes \cite{kroy2005, fischer2008} has been demonstrated experimentally \cite{groh2009a}.

Minimal models are continuum models: They deal with the overall shape of the dunes neglecting the particulate nature of these granular systems. However, the dynamics of dunes is determined by the transport mechanism of the individual grains of sand \cite{bagnold1941}. The aeolian sand transport consists of two modes of transport: \emph{reptation} in the low energy regime and \emph{saltation} in the high energy regime \cite{andreotti2004}. The saltating grains are carried with the wind and have flight lengths of thousands of grain diameters. The paths of the reptating grains are much shorter. Their motion is initiated by impacts of the saltating grains onto the granular bed \cite{ungar1987}. Experiments in wind-tunnels with sand traps \cite{dong2002, ni2002, zhou2002} and with particle tracking methods \cite{nishimura2000, yang2007, creyssels2009} give insight into the aeolian grain transport on the level of particle dynamics.

Neglecting the possibility of suspension at high shear velocities, the bed-load transport in water can also be separated in two energy regimes: saltation and surface creep \cite{bagnold1973, collinson1989, simons1992, osanloo2008}. The saltating particles have much shorter flight lengths than in air and, in contrast to the reptating particles, the creeping particles are directly dragged by the fluid and stay always in contact with the bed surface. Particle tracking experiments in flow channels with denser fluids than air, like water or silicon oil, have been performed to investigate the erosion, the transport, and the deposition of grains of sand on a granular bed \cite{anecy2002, anecy2003, charru2004, lobkovsky2008, mouilleron2009}.

The particle dynamics at the surface transfers itself into the bed. The packing density of the granular bed decreases from the inner layers towards the outer fluidized layer, which has a thickness of a few grain diameters in a laminar flow \cite{lobkovsky2008}. The observation of segregation effects indicate that the sediment in the sand bed is mixed during the process of the sand transport and the associated ripple formation \cite{rousseaux2004}. Besides the global granular transport, the transition between creeping and saltating particles is also a matter of interest \cite{dong2002, anecy2002, anecy2003, wang2004, osanloo2008}.

We investigate a migrating isolated dune on a plane surface reminiscent of a barchan dune in the desert. Our experiment is supposed to reveal the details of the particle transport above, on, and inside this solitary dune. Together with the characterization of the driving water flow field, the results should encourage future particulate models and simulations \cite{narteau2009}. To get access to the dynamics of the individual grains, we create an experimental realization of a minimal dune model --- a cartoon dune.

\begin{figure}
\begin{center}
\includegraphics[width = 0.75\linewidth]{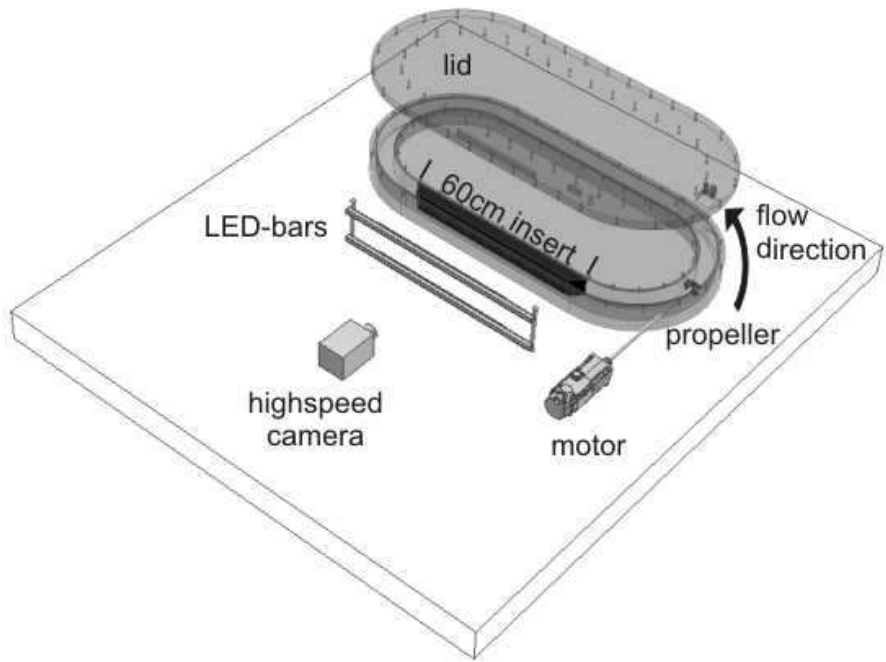}
\caption{\label{fig1}Sketch of the experimental setup.}
\end{center}
\end{figure}

A sketch of our experimental setup is shown in Fig.\,\ref{fig1}. The main part consists of a closed flow channel machined from perspex. The height of the channel amounts to $H = 60$\,mm, and its width is 50\,mm. The length of the straight section is 600\,mm, the curves have an outer diameter of 500\,mm and an inner diameter of 400\,mm. The channel is filled up with distilled water. The flow is generated by a propeller with a diameter of 45\,mm, which is installed in the curve following the section of measurements. The propeller is driven by a motor with a shaft which is placed in the middle of the channel profile. The flow direction is counter-clockwise on top view. We narrow the section of measurements to create quasi two-dimensional conditions, comparable to a Hele-Shaw cell \cite{heleshaw1898}. For that purpose we use a black plastic insert, which constricts the channel to a width $W = 3$\,mm. This leads to an aspect ratio of $H / W = 20$.

For the calculation of the Reynolds number \textrm{Re} in the 3\,mm wide gap we measure the flow velocity $\vec{u}$ with an ultrasonic Doppler velocimeter (UDV) manufactured by \emph{Signal Processing SA}. This device measures the vertical velocity profiles $\vec{u} (y)$. From these profiles we extract the mean horizontal velocity $u_{\textnormal{\scriptsize{mean}}}$ by averaging along the central 80 percent of the profiles. The kinematic viscosity of water at experimental temperature ($20.5 \pm 0.3$\,$^{\circ}$C) is $\nu \approx 1$\,mm$^{2}$/s. The flow velocity is kept at $u_{\textnormal{\scriptsize{mean}}} = 0.65$\,m/s, which corresponds to $\textrm{Re}= u_{\textnormal{\scriptsize{mean}}} H / \nu = 39000$.
Instead of natural sand we use spherical white glass beads (\emph{SiLibeads}), which have a radius of $r = 1.00 \pm 0.01$\,mm and a density of $\rho = 2.51$\,g/cm$^{3}$. For the model dune, the overall mass of the beads amounts to $m = 6.5$\,g, which corresponds to 629 particles.

\begin{figure}
\begin{center}
\includegraphics[width = 0.6\linewidth]{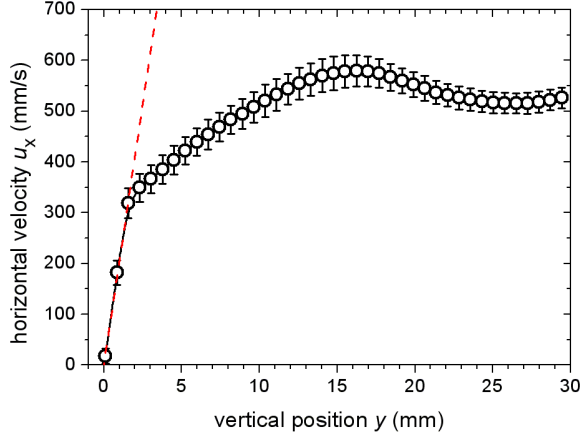}
\caption{\label{fig2}Velocity profile $u_{\textnormal{\scriptsize{x}}}(y)$ of the water flow at the threshold of incipient motion. The dashed red line is a linear fit to the three lowest data points yielding the shear velocity $u^{*}$.}
\end{center}
\end{figure}

To characterize the particle transport in our experiment, we investigate the threshold of incipient motion for the used glass beads. Therefore, we prepare a flat granular bed and determine the shear velocity
\begin{displaymath}
u^{*} = \sqrt{ \left. \nu \frac{ \textnormal{d$u_{\textnormal{\scriptsize{x}}}$} } {\textnormal{d$y$}}  \right| _{\textnormal{\footnotesize{surface}}} }
\end{displaymath}
near the surface. The corresponding profile $u_{\textnormal{\scriptsize{x}}}(y)$ and the slope within the laminar boundary layer
are shown in Fig.\,\ref{fig2}. We observed that particle motion sets in at the critical shear velocity $u^{*}_{\textnormal{\footnotesize{c}}} = 14$\,mm/s. This number is used for the calculation of the critical particle Reynolds number
\begin{displaymath}
\textrm{Re}_{\textnormal{\footnotesize{c}}} = \frac{u^{*}_{\textnormal{\footnotesize{c}}}2r}{\nu}
\end{displaymath}
and the critical Shields parameter
\begin{displaymath}
\Theta_{\textnormal{\footnotesize{c}}} = \frac{u^{*2}_{\textnormal{\footnotesize{c}}}}{g(s-1)2r}
\end{displaymath}
with the acceleration of gravity $g = 9.81$\,m/s$^{2}$, the density of water $\rho_{\textnormal{\footnotesize{f}}} = 1$\,g/cm$^{3}$, and the specific density $s = \rho / \rho_{\textnormal{\footnotesize{f}}}$. The resulting $\textrm{Re}_{\textnormal{\footnotesize{c}}} = 28$ is about six times larger than the typical value for aeolian dunes $\textrm{Re}_{\textnormal{\footnotesize{c, air}}} = 5$, whereas $\Theta_{\textnormal{\footnotesize{c}}} = 0.007$ is comparable to its aeolian counterpart $\Theta_{\textnormal{\footnotesize{c, air}}} = 0.010$ \cite{groh2009a}.

The model dune is observed with a high-speed camera (\emph{IDT}, MotionScope M3) which is placed in front of the straight part of the channel. The camera is set to a resolution of $1280 \times 256$ pixels at 2000 frames per second. For a sufficiently bright illumination we use two bars of light-emitting diodes (LED-bars), as shown in Fig.\,\ref{fig1}. They lighten the section of measurements from below and from above to obtain a homogenous brightness.

\begin{figure}
\begin{center}
\includegraphics[width = 0.75\linewidth]{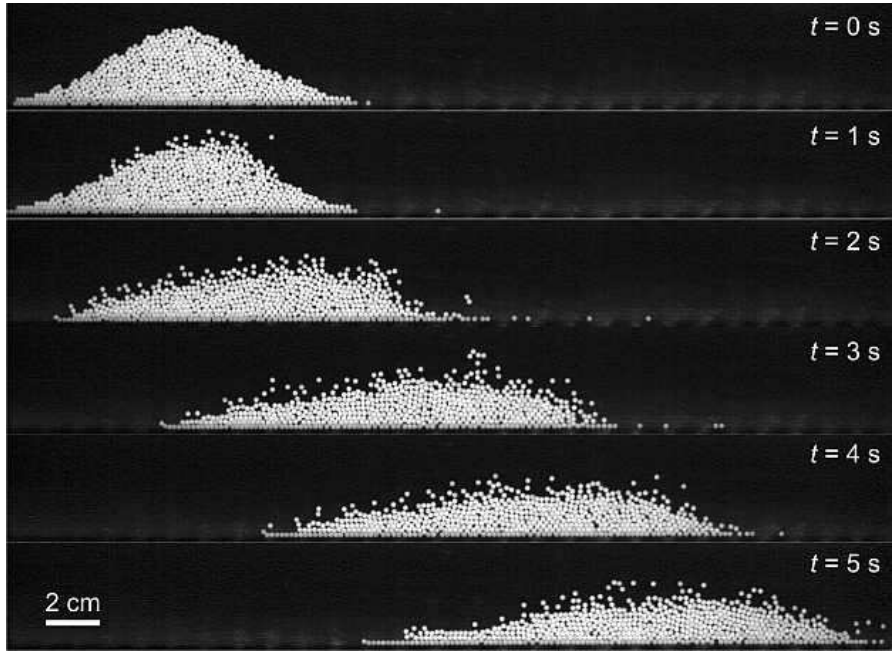}
\caption{\label{fig3}Six time sequenced snapshots showing side views of a developing
barchan dune slice composed of white glass beads.}
\end{center}
\end{figure}

The experimental procedure is as follows: After the flow tube is filled with distilled water, a funnel with a 3\,mm long slit is used to pour glass beads into the channel. The experiment starts with the triangular heap shown in Fig.\,\ref{fig3} at $t = 0$\,s. The five subsequent snapshots show the temporal evolution of the initial triangular heap towards an asymmetric heap moving downstream. After $t = 3$\,s the heap reaches the characteristic steady-state shape known from central slices of barchan dunes along their migration direction \cite{sauermann2000}. In the images the glass beads appear as bright disks in front of the dark background. For their localization we use the Circle Hough Transformation as described in detail in Refs.\,\cite{duda1972,kimme1975}. With this image-processing technique all bead positions in every recorded frame are found. The high-speed camera allows particle tracking for each individual grain, which makes it possible to study the particle dynamics of dunes and the mixing process within dunes.

\begin{figure}
\begin{center}
\includegraphics[width = 0.75\linewidth]{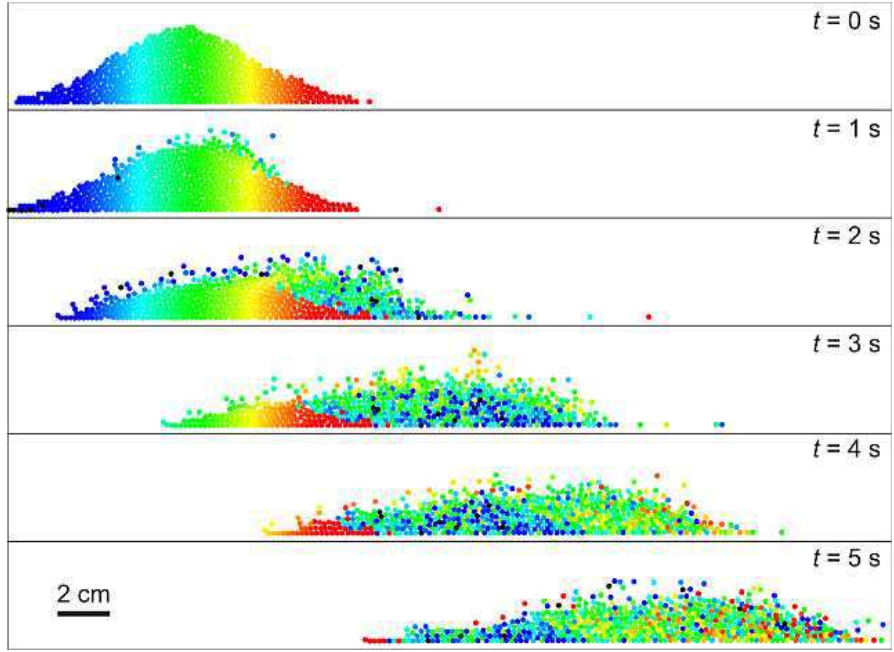}
\caption{\label{fig4}The particles detected in Fig.\,\ref{fig3} are drawn as disks and color coded with respect to their horizontal positions at $t = 0$\,s (see animated movie1).}
\end{center}
\end{figure}

To visualize the mixing we provide Fig.\,\ref{fig4}: At the beginning all detected beads of the data set are colored with a continuous color gradient from the left to the right, as the snapshot at $t = 0$\,s  shows. As time evolves the traceable beads keep their initial color, the non-traceable ones are painted black. As the flow sets in, the grains at the surface of the upstream-side are carried away and fall onto the downstream-side, where they roll further downwards. In nature this side is called the slipface. From the color code, it can be seen that the inner beads rest in place. The first cycle of the mixing is passed after $t = 5$\,s, when again most of the blue circles are at the windward side.

\begin{figure}
\begin{center}
\includegraphics[width = \linewidth]{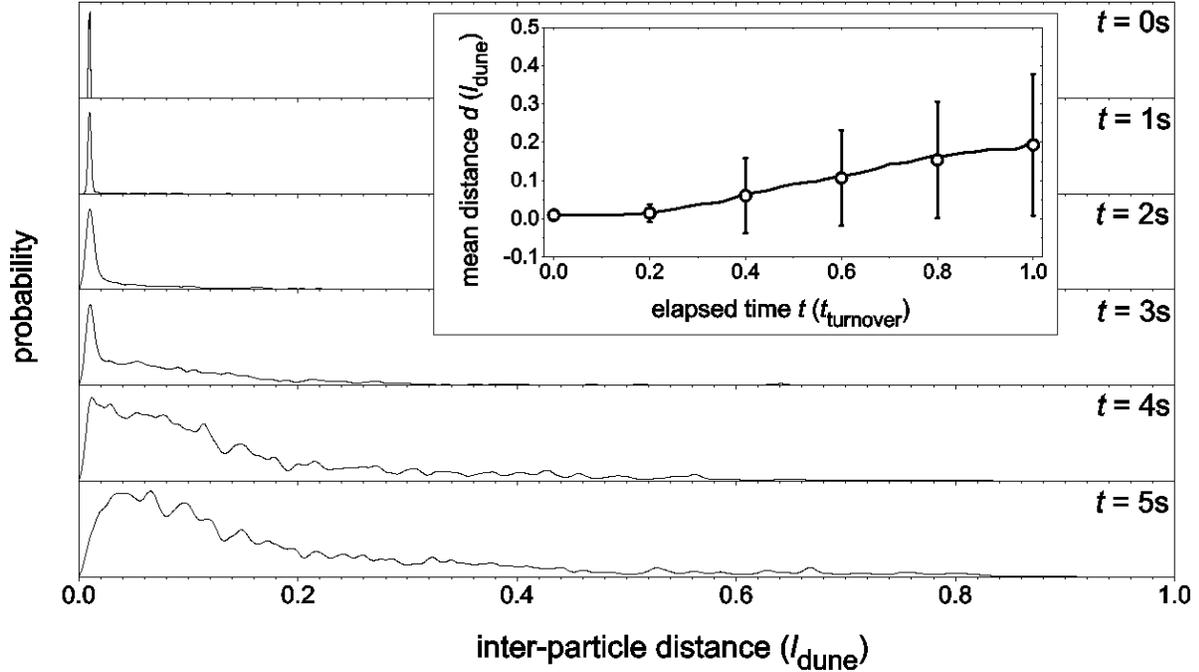}
\caption{\label{fig5}Six histograms corresponding to Fig.\,\ref{fig4} showing the temporal evolution of the distribution of distances $d_{i,j}$. The probability distributions are individually scaled with respect to their maxima. The insert shows the temporal evolution of the mean distance $d$. The six data points correspond to the histograms and their standard deviations are represented by the error bars.}
\end{center}
\end{figure}

For a quantitative measure of the mixing process we investigate the temporal evolution of the distances $d_{i,j}$ between those particles $i$ and $j$ which are initially in contact. Six representative histograms are plotted in Fig.\,\ref{fig5}. The initially sharp distribution broadens with time, and after $t = 5$\,s the distribution spans the complete basis dune length $l_{\textnormal{\scriptsize{dune}}}$. This basis length is determined by the bulk part of the dune as defined later (see Fig.\,\ref{fig14}(a)) and turns out to be $l_{\textnormal{\scriptsize{dune}}} = 190$\,mm. The growth of the mean distance $d = <d_{i,j}>$ and its standard deviation is shown in the insert of Fig.\,\ref{fig5}. It indicates that within the first turnover time $t_{\textnormal{\scriptsize{turnover}}}$ of about 5\,s the mean distance between particles initially in contact reaches a value of approximately one quarter of the dune basis length.

\begin{figure}
\begin{center}
\includegraphics[width = 0.6\linewidth]{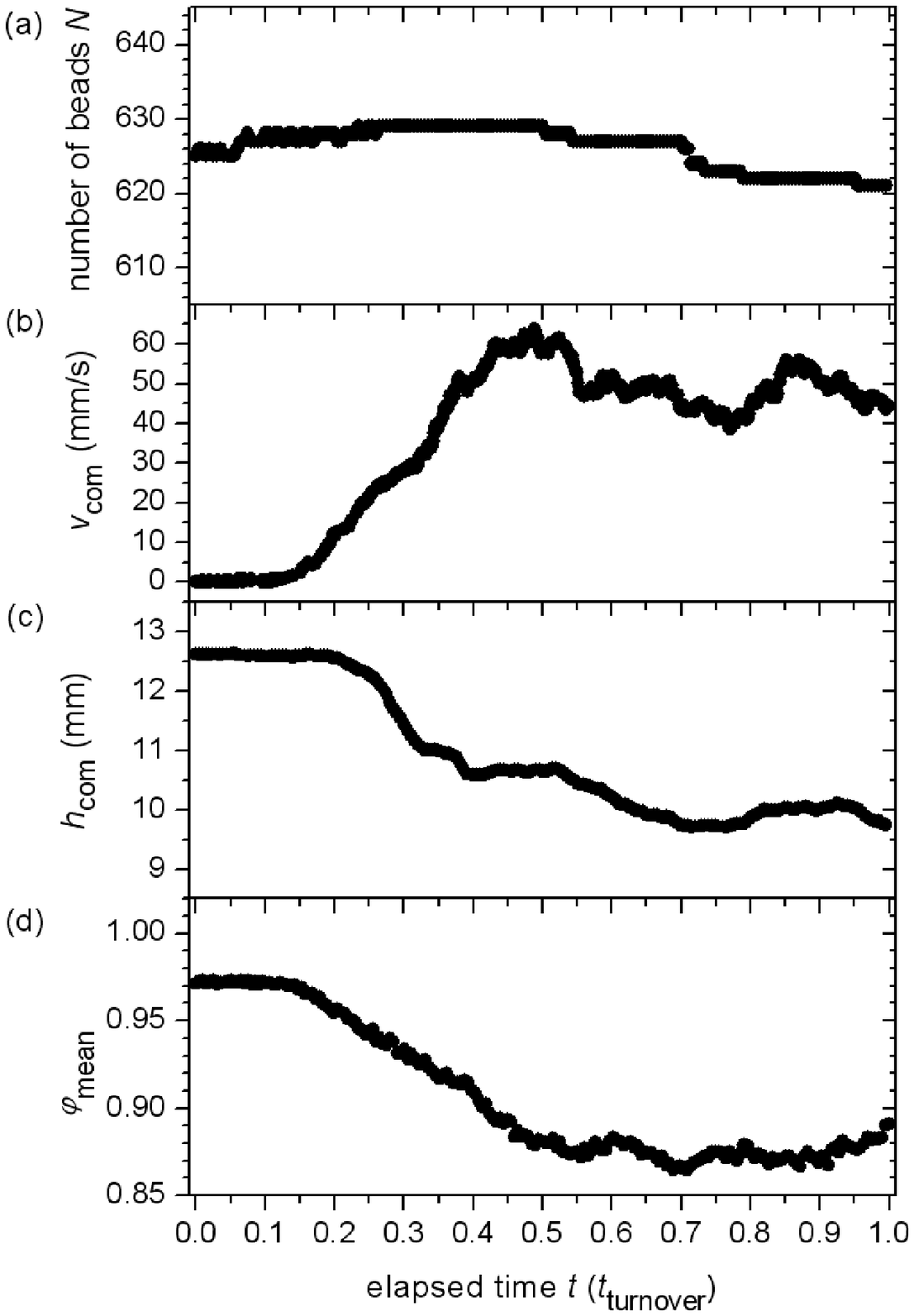}
\caption{\label{fig6}The time evolution of (a) the number of beads $N$, (b) the velocity of the center of mass of the dune $v_{\textnormal{\scriptsize{com}}}$, (c) the height of the center of mass $h_{\textnormal{\scriptsize{com}}}$, and (d) the temporal evolution of the mean packing fraction $\varphi_{\textnormal{\scriptsize{mean}}}$ obtained from the Voronoi method. Every 50th data point is plotted.}
\end{center}
\end{figure}

Moreover, once all particle positions are known, other quantities describing the dynamics of the dune can be extracted.  As shown in Fig.\,\ref{fig6}(a) the number of glass beads $N$ is slightly fluctuating within one percent. This is primarily due to particles flowing in and out of the frame, and partly caused by detecting failures. Although only the particles from the surface are in motion, the center of mass of the dune migrates downstream with a velocity $v_{\textnormal{\scriptsize{com}}}$. From the positions of all beads in the frame, both the center of mass and its velocity are determined. It can be seen in Fig.\,\ref{fig6}(b) that $v_{\textnormal{\scriptsize{com}}}$ takes a constant value after $t = 2$\,s. This coincides with the relaxation towards the steady-state shape, which is demonstrated by the constant height of the center of mass $h_{\textnormal{\scriptsize{com}}}$ in Fig.\,\ref{fig6}(c).

Alternatively, the relaxation process can be characterized by measuring the area packing fraction $\varphi_{\textnormal{\scriptsize{a}}}$ within the core of the dune, which is composed of the inner static particles. For that purpose we calculate the Voronoi tessellation of each frame \cite{qhull}. The packing fraction is obtained as the ratio between the cross sectional area of a spherical glass bead and the corresponding area of its Voronoi cell. In order to characterize only the core beads, we exclude those beads with $\varphi_{\textnormal{\scriptsize{a}}} < 2/3$\,$\varphi_{\textnormal{\scriptsize{max}}}$. The maximal packing fraction $\varphi_{\textnormal{\scriptsize{max}}}$ can be larger than one, because an overlap of particles is possible in the two-dimensional projection of a three-dimensional packing. In our narrow channel of thickness $3r$ the maximal packing fraction $\varphi_{\textnormal{\scriptsize{max}}} = \pi/3$ is obtained by assuming a two-dimensional square lattice with a lattice constant of $2 \sqrt{3} r$. The crystal basis consists of two beads in contact forming an angle of $30^{\circ}$ with respect to the plane of this lattice.

\begin{figure}
\begin{center}
\includegraphics[width = 0.75\linewidth]{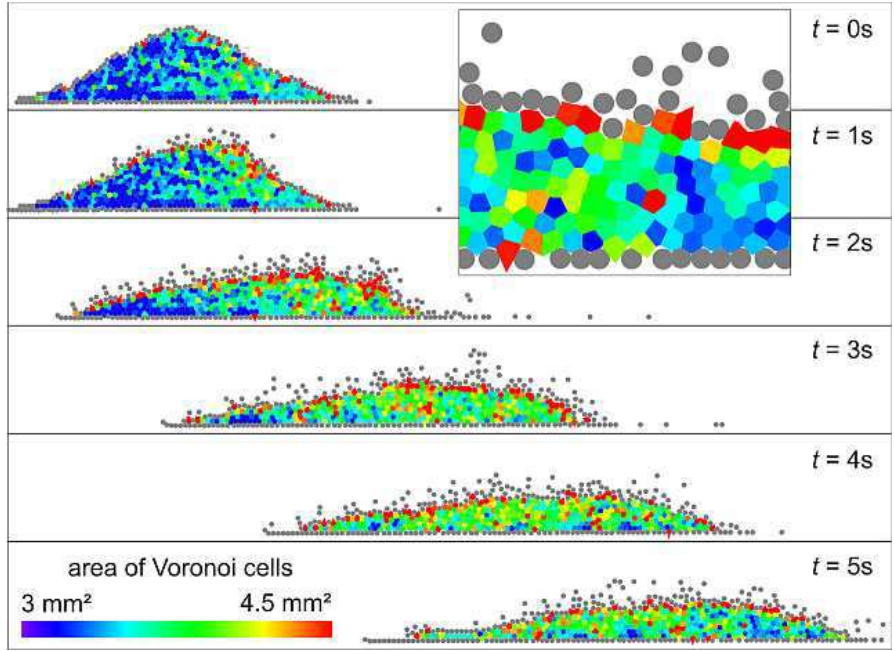}
\caption{\label{fig7}The particles detected in Fig.\,\ref{fig3} are drawn as grey disks. Overlayed are the Voronoi cells of the inner particles, being color coded with respect to their size. The insert shows a detail of the snapshot at $t = 5$\,s.}
\end{center}
\end{figure}

In Fig.\,\ref{fig7} the resulting Voronoi cells of the six snapshots of Fig.\,\ref{fig3} are overlayed to all detected beads drawn as grey disks. The cells are color coded with respect to their area. The loss of the blue cells indicates that the packing fraction decreases during the time of measurements. This is quantitatively demonstrated by Fig.\,\ref{fig6}(d), where the temporal evolution of the spatial mean packing fraction $\varphi_{\textnormal{\scriptsize{mean}}}$ corresponding to these Voronoi cells is plotted. In addition to the relaxation of $v_{\textnormal{\scriptsize{com}}}$ and $h_{\textnormal{\scriptsize{com}}}$, $\varphi_{\textnormal{\scriptsize{mean}}}$ also takes a constant value after $t = 3$\,s.

The investigation of the transient demonstrates that the initial triangular heap evolves to a steady state with constant mass, shape, velocity, and packing fraction after about one turnover time $t_{\textnormal{\scriptsize{turnover}}}$ has elapsed. Within that time the mean distance between particles initially in contact reaches a value of approximately one quarter of the dune length $l_{\textnormal{\scriptsize{dune}}}$.

The presented Voronoi method has the advantage to reveal the temporal evolution of the packing fraction, but has the disadvantage to be operational only within a small fraction near the core of the dune. In order to get meaningful results also in the surrounding area of the dune, we obtain the local area packing fraction from the possibility for each pixel to be shaded by a particle during a certain time. The relation between $\varphi_{\textnormal{\scriptsize{a}}}$ and the volume packing fraction $\varphi_{\textnormal{\scriptsize{v}}}$ is
\begin{displaymath}
\varphi_{\textnormal{\scriptsize{v}}} = \frac{4 N \pi r^{3}}{3 A W} = \frac{4 r}{3 W} \varphi_{\textnormal{\scriptsize{a}}} \qquad ,
\end{displaymath}
where $N$ is the number of particles in a certain area $A$. In our geometry $r/W = 1/3$ and thus $\varphi_{\textnormal{\scriptsize{v}}} = 4/9$\,$\varphi_{\textnormal{\scriptsize{a}}}$. This yields for the maximum value $4/9 \cdot \pi/3 \approx 0.47$.

\begin{figure}
\begin{center}
\includegraphics[width = 0.75\linewidth]{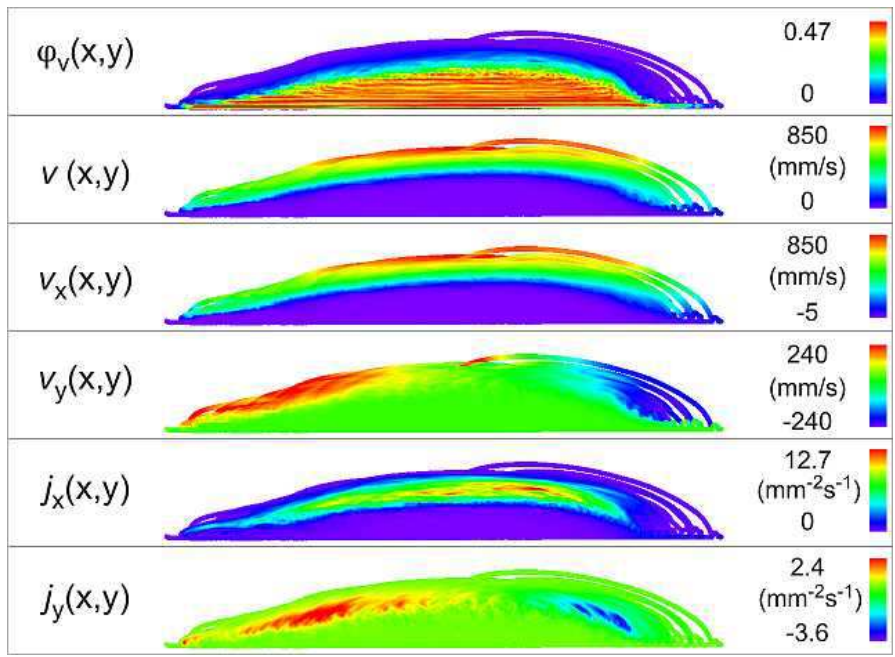}
\caption{\label{fig8}The time averaged fields of the measurement shown in Fig.\,\ref{fig3} between $t = 4$\,s and $t = 5$\,s. The width corresponds to 33.7\,cm in the experiment, and the height of each panel to 4.1\,cm. The numbers on the right-hand side of each panel correspond to the minimal and maximal value of the color code.}
\end{center}
\end{figure}

The experimentally obtained field $\varphi_{\textnormal{\scriptsize{v}}} (x,y)$ is shown in Fig.\,\ref{fig8}. Here, the time average is taken over one second, which corresponds to 2000 single frames. For technical reasons the frames are shifted according to the center of mass of the dunes.

The mean velocity field of the particles is calculated similar to $\varphi_{\textnormal{\scriptsize{v}}} (x,y)$. Its absolute value $v (x,y)$, its x-component $v_{\textnormal{\scriptsize{x}}} (x,y)$, and its y-component $v_{\textnormal{\scriptsize{y}}} (x,y)$ are plotted in Fig.\,\ref{fig8}. The field $v (x,y)$ shows that the velocity increases with increasing height above the dune surface. This indicates that the grain movement changes from creeping to saltation --- i.e. freely swimming in the water stream. This behavior is also reflected by $v_{\textnormal{\scriptsize{x}}} (x,y)$. Notably, $v_{\textnormal{\scriptsize{x}}} (x,y)$ has almost no negative values, indicating that the recirculation bubble at the downstream side \cite{ayrton1910} is not strong enough to push the sinking beads backwards. The field $v_{\textnormal{\scriptsize{y}}} (x,y)$ visualizes the erosion and deposition process: at the upstream side the particles are lifted and above the slipface they rapidly fall down.

\begin{figure}
\centering \subfigure[\label{fig9}]{\includegraphics[width = 0.75\linewidth]{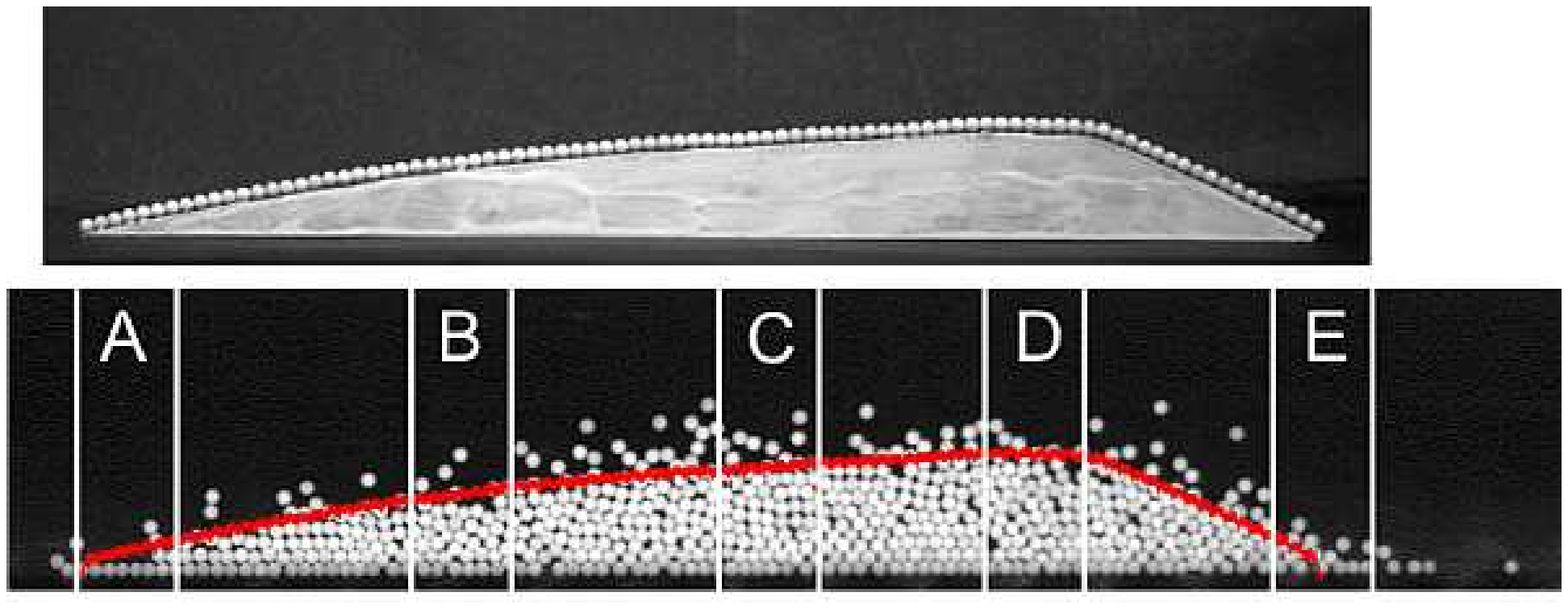}} \subfigure[\label{fig10}]{\includegraphics[width = 0.9\linewidth]{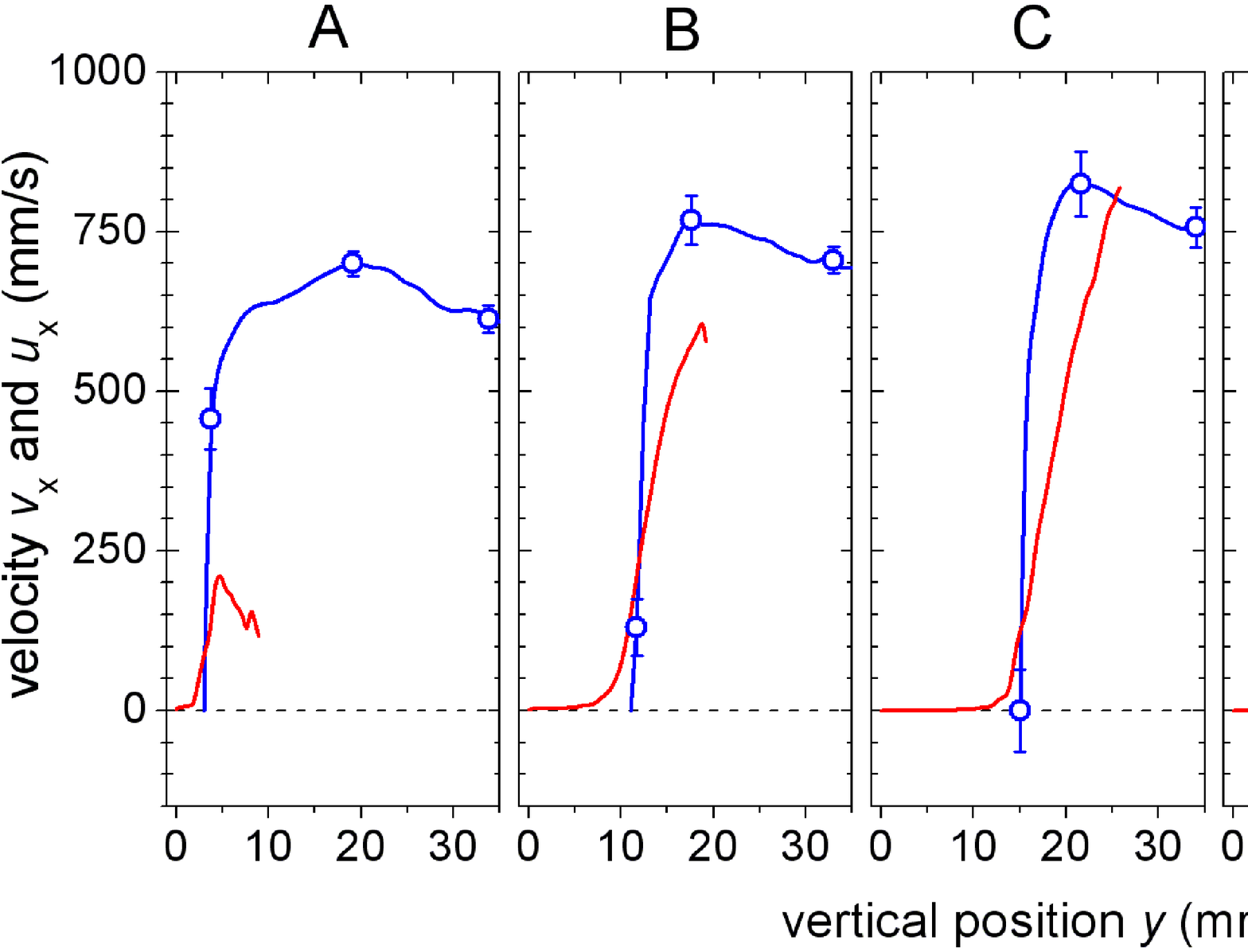}} \caption{Panel (a) shows the plastic dummy decorated with glass beads and a detail of the snapshot at $t = 4$\,s in Fig.\,\ref{fig3}. The red line indicates the profile of the plastic dummy used for the determination of the water flow field. The five areas tagged with white lines and capitals correspond to panel (b), which provides the comparison between the vertical velocity profiles $v_{\textnormal{\scriptsize{x}}}(y)$ of grains (red lines) and $u_{\textnormal{\scriptsize{x}}}(y)$ of water flow (blue lines) in the selected areas. Each profile consists of 100 measured data points of which a few representatives are plotted with error bars.}
\end{figure}

The motion of the individual particles results from the interplay with the driving water flow. To illuminate this interaction we need the velocity field of the water flow, which can be measured with the UDV. However, the UDV needs a couple of seconds to acquire a single velocity profile. This is too long for recording the velocity field within the time the dune needs to propagate along the section of measurements. We evade this problem by measuring the flow field around a steady dune made of plastic instead of the moving dune. The contour of this dummy is obtained from the snapshot at $t = 4$\,s in Fig.\,\ref{fig3} by drawing a smoothed envelope around the densely packed bulk of the dune. The result is shown as a red line in Fig.\,\ref{fig9}. The dummy is decorated with glass beads to simulate the surface roughness, which determines the size of the laminar boundary layer.

We define five characteristic areas (A - E) as shown in Fig.\,\ref{fig9} to compare the vertical velocity profiles of the grains and the water flow in Fig.\,\ref{fig10}. The vertical grain velocity profiles $v_{\textnormal{\scriptsize{x}}} (y)$ are extracted from $v_{\textnormal{\scriptsize{x}}} (x,y)$ by averaging horizontally in the corresponding areas. The vertical water velocity profiles $u_{\textnormal{\scriptsize{x}}}(y)$ are recorded by crossing two ultrasonic beams above the dune surface to get an averaged data set within each area.

The highest measured velocities are found near the crest of the dune. For the water flow this is a manifestation of the incompressibility condition in our closed channel. It is notable that in the areas A to D the particle velocity is finite at such positions $y$ where the water velocity is zero. This can be explained by the fact that the flow velocities are measured with the plastic dummy, which fails to take into account the motion of the particles on the surface. In area E the water flow velocity changes its sign at lower heights, which is an indication of the recirculation bubble. In contrast, the grain motion does not follow the water flow in the wake of the dune. In all areas the velocity gradients of the water profiles are steeper than the ones of the beads. This might result from the fact, that the particles cannot follow the water flow immediately due to inertia. Notably, the maximal velocities of both, water and beads, is one order of magnitude greater than the migration velocity of the dune measured as the center-of-mass motion. This is explained by the particle size compared to the dune size.

\begin{figure}
\begin{center}
\includegraphics[width = 0.6\linewidth]{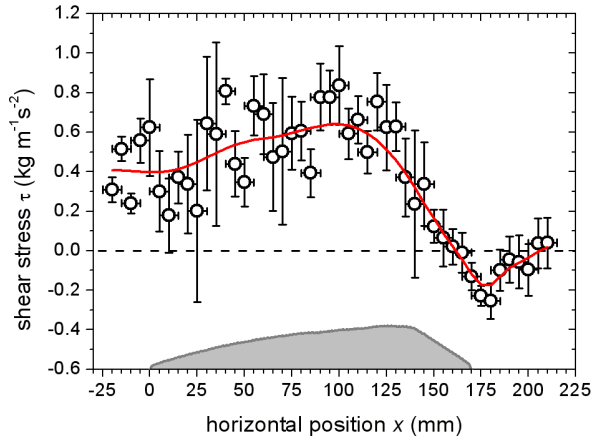}
\caption{\label{fig11}Shear stress of the water flow at the dune surface. A smoothing of the data points yields the red line, which is supposed to guide the eye. The grey shaded area symbolizes the location and the shape of the plastic dummy.}
\end{center}
\end{figure}

From the velocity gradient of the water flow near the surface of the dune(see Fig.\,\ref{fig2}), we calculate the shear stress of the water flow
\begin{displaymath}
\tau = \left. \eta \frac{ \textnormal{d$u_{\textnormal{\scriptsize{x}}}$} } {\textnormal{d$y$}}  \right| _{\textnormal{\scriptsize{surface}}}
\end{displaymath}
assuming $\eta = 10^{-3}$\,Pas for water at $20$\,$^{\circ}$C. Here, the additional turbulent shear stress is neglected, because near fixed walls in the area of the viscous boundary layer the eddy viscosity goes to zero \cite{oertel2009}. The resulting values for $\tau$ as shown Fig.\,\ref{fig11} quantify the slope of the water profiles near the surface of the dune and give a measure for the local drag force. The shear stress increases along the upstream side of the dune and reaches a maximum value afore the crest. Above the slipface $\tau$ decreases and even reverses sign in the region of the recirculation bubble and relaxes further downstream. This qualitative behavior corresponds to the analytical description of the wind shear stress used in the theoretical model by Kroy \etal \cite{kroy2002a, kroy2002b}.

\begin{figure}
\centering \subfigure[\label{fig12}]{\includegraphics[width = 0.9\linewidth]{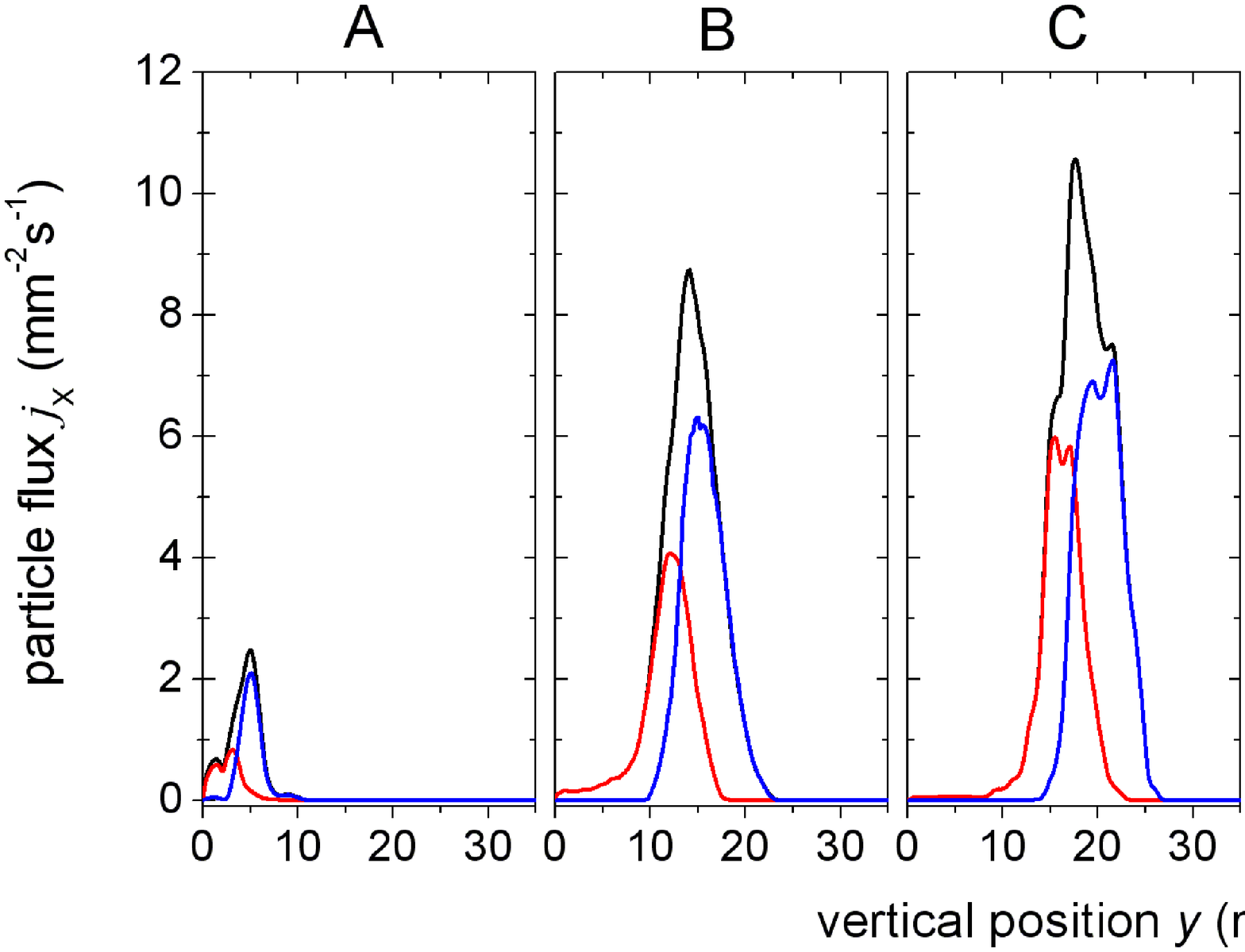}} \subfigure[\label{fig13}]{\includegraphics[width = 0.9\linewidth]{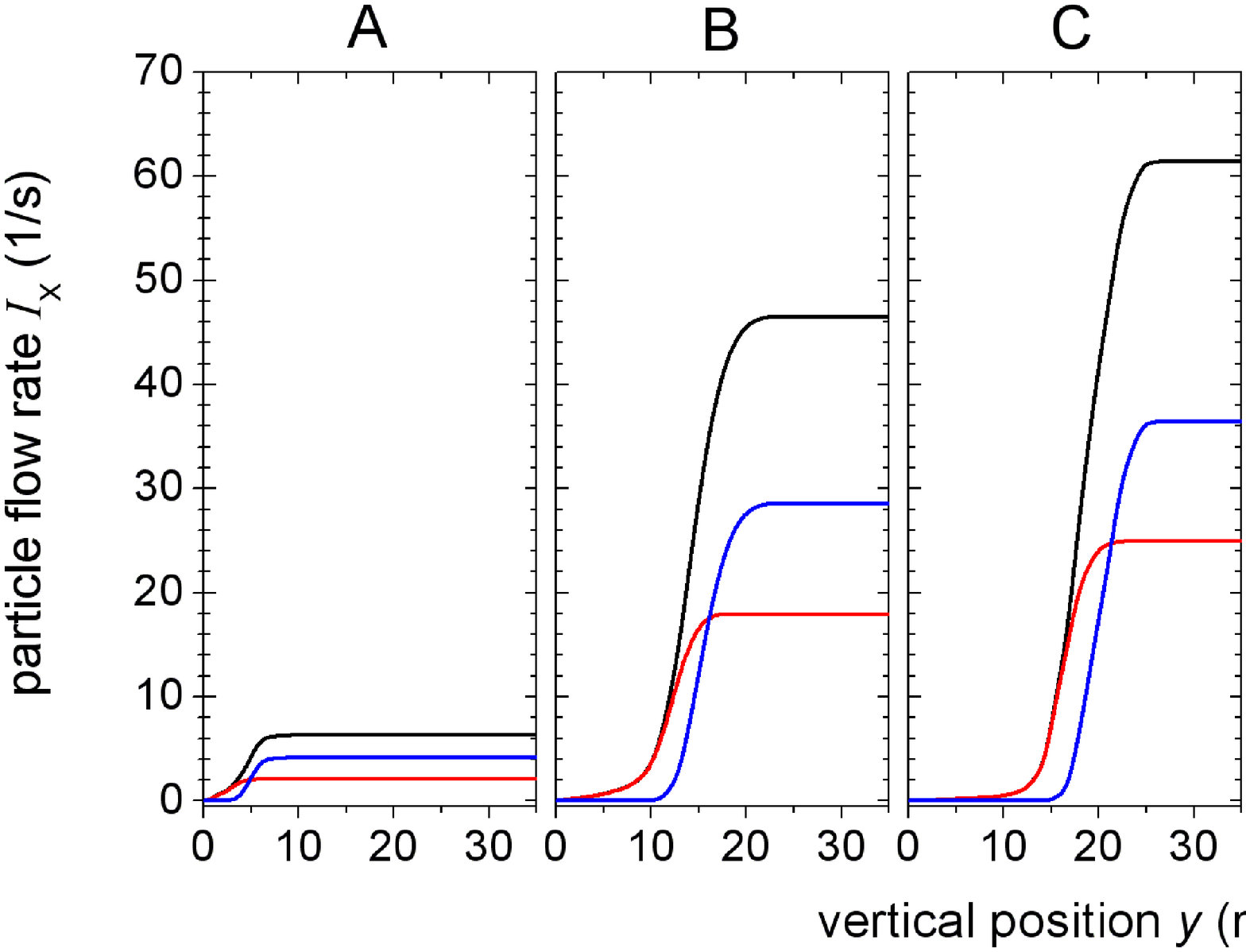}} \caption{Panel (a) shows the particle flux $\emph{\textrm{j}}_{\textnormal{\scriptsize{x}}} (y)$ and panel (b) shows the particle flow rate $\emph{\textrm{I}}_{\textnormal{\scriptsize{x}}} (y)$ for all beads (black lines), beads with contacts (red lines), and freely moving beads (blue lines). The data sets are obtained from the corresponding five areas in Fig.\,\ref{fig9}.}
\end{figure}

The resulting particle fluxes $\emph{\textrm{j}}_{\textnormal{\scriptsize{x}}} (x,y)$ and $\emph{\textrm{j}}_{\textnormal{\scriptsize{y}}} (x,y)$ are shown in Fig.\,\ref{fig8}. They are the product of the particle density $\varphi_{\textnormal{\scriptsize{v}}} (x,y)$/$V_{\textnormal{\scriptsize{bead}}}$ times $v_{\textnormal{\scriptsize{x}}} (x,y)$ or $v_{\textnormal{\scriptsize{y}}} (x,y)$, respectively, where $V_{\textnormal{\scriptsize{bead}}} = 4\pi/3$\,mm$^{3}$ is the volume of a glass bead. In analogy to Fig.\,\ref{fig10} we extract the averaged vertical flux profiles of the horizontal flux $\emph{\textrm{j}}_{\textnormal{\scriptsize{x}}} (y)$ as presented in Fig.\,\ref{fig12}. From these profiles the particle flow rates as a function of the vertical position
\begin{displaymath}
\emph{\textrm{I}}_{\textnormal{\scriptsize{x}}} (y) = \int_{0}^{W} \int_{0}^{y} \emph{\textrm{j}}_{\textnormal{\scriptsize{x}}} (y') \textnormal{d$y'$}\textnormal{d$z$}
\end{displaymath}
are extracted by integrating $\emph{\textrm{j}}_{\textnormal{\scriptsize{x}}} (y)$ with the vertical position $y$ and the width $W$ of the flow channel. The resulting curves are plotted in Fig.\,\ref{fig13}.

\begin{figure}
\begin{center}
\includegraphics[width = 0.8\linewidth]{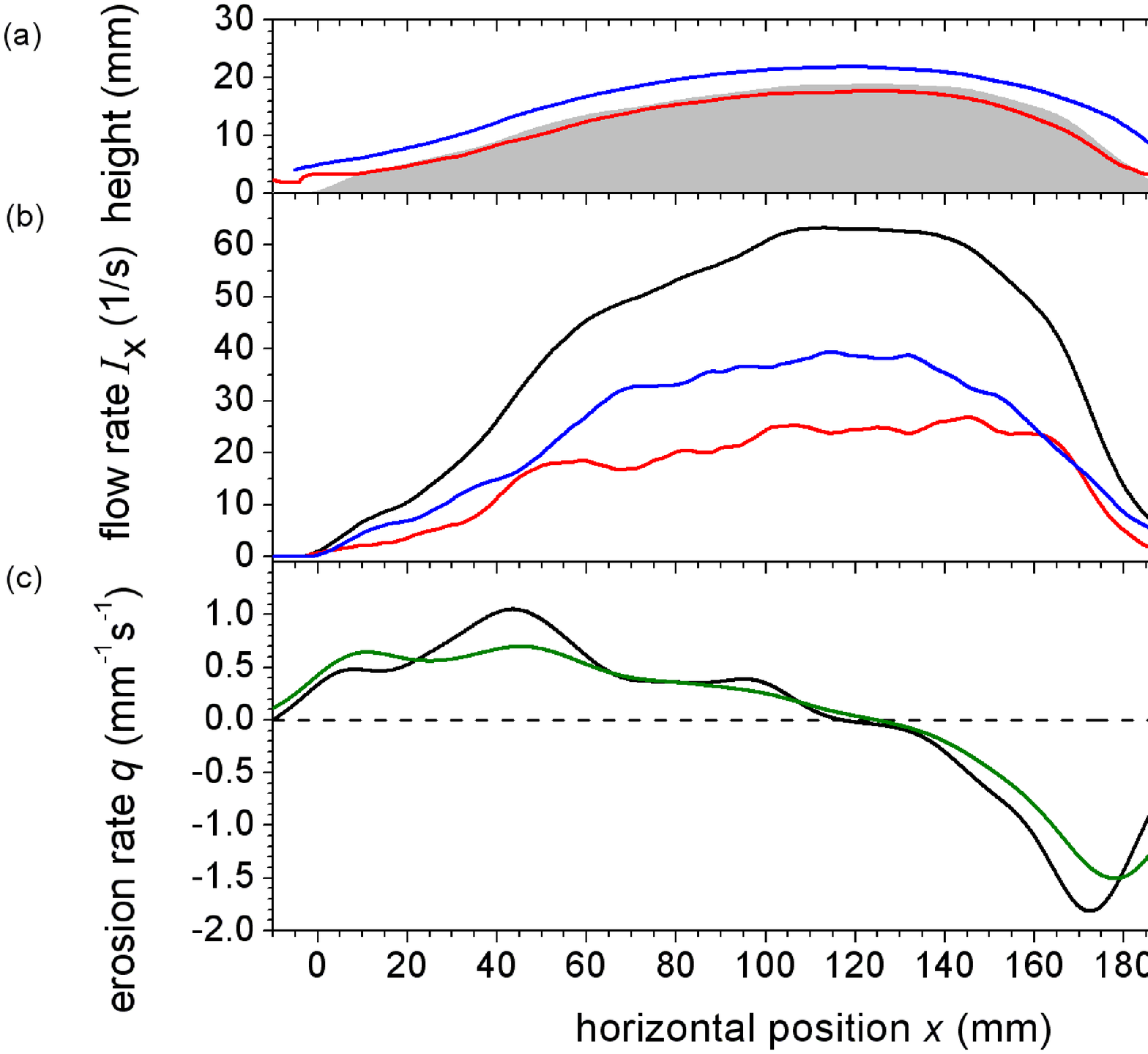}
\caption{\label{fig14}(a) Spatial profiles of the averaged vertical positions of creeping (red) and saltating (blue) particles. The grey shaded area symbolizes the bulk of the dune. (b) Spatial profiles of flow rates $I_{\textnormal{\scriptsize{x,creep}}}$ for creeping (red) and $I_{\textnormal{\scriptsize{x,salt}}}$ for saltating (blue) particles. The black line indicates the sum $I_{\textnormal{\scriptsize{x,total}}}$ of both types. (c) Spatial profile of the local erosion rate $q$ (black line) and the local slope of the dune surface (green line).}
\end{center}
\end{figure}

To specify the nature of the particle transport, we distinguish between beads which have contacts to other beads forming the bulk shown as gray shaded area in Fig.\,\ref{fig14}(a), and beads which move freely in the water stream without contact to the bulk. The contact criteria is fulfilled, when the centers of two beads take a distance shorter than $2.1 r$. The beads within the bulk are composed of resting beads and creeping beads, but only the creeping beads contribute to the flux density. The motion of the free flying particles above the surface is named saltation. It can be seen that at every position along the dune the average vertical positions of creeping particles is near the surface $h(x)$ and the average vertical positions of saltating particles is located two to three particle diameters above. The vertical positions are determined as the first moment of the vertical particle flux distributions. The distributions are calculated with a spatial running average along the dune with a fixed width of 13\,mm corresponding to the areas in Fig.\,\ref{fig8}.

We define the values of the spatially averaged integral particle flow rate $I_{\textnormal{\scriptsize{x,creep}}} = I_{\textnormal{\scriptsize{x}}}(y = H)$ for creeping particles. $I_{\textnormal{\scriptsize{x,salt}}}$ for saltating particles is defined likewise. The total particle flow rate results as $I_{\textnormal{\scriptsize{x,total}}} = I_{\textnormal{\scriptsize{x,creep}}} + I_{\textnormal{\scriptsize{x,salt}}}$. All three flow rates are shown as a function of the horizontal position in Fig.\,\ref{fig14}(b). The spatial profiles of the separated flow rates show, that the saltating beads contribute a larger part of the dune motion than the creeping beads. Both flow rates reach their maximum values near the crest of the dune, as it is known from natural dunes in the desert \cite{sauermann2001, sauermann2003}.

The overall erosion rate
\begin{displaymath}
q = \frac{\textnormal{d}^{2}N}{\textnormal{d}t \textnormal{d}x}
\end{displaymath}
can be obtained as the spatial derivative of the total flow rate $\textnormal{d}I_{\textnormal{\scriptsize{x,total}}} / \textnormal{d}x$. The erosion rate as shown in Fig.\,\ref{fig14}(c) matches qualitatively with the local slope of the dune surface $\textnormal{d}h / \textnormal{d}x$. If $q > 0$ the beads are entrained into the water stream and will strengthen the total flow rate. If $q < 0$ the beads are deposited on the surface of the dune and the value of $I_{\textnormal{\scriptsize{x,total}}}$ will decrease. Along the upstream side $q$ is greater than zero and becomes negative behind the crest with a maximal deposition onto the slipface. The plateau near the crest indicates a constant flow rate. These observations correspond to the color code of $\emph{\textrm{j}}_{\textnormal{\scriptsize{y}}} (x,y)$ in Fig.\,\ref{fig8}. Notably, the position of the maximal deposition matches with the position of the maximal negative shear stress in Fig.\,\ref{fig11}.

\begin{figure}
\begin{center}
\includegraphics[width = 0.6\linewidth]{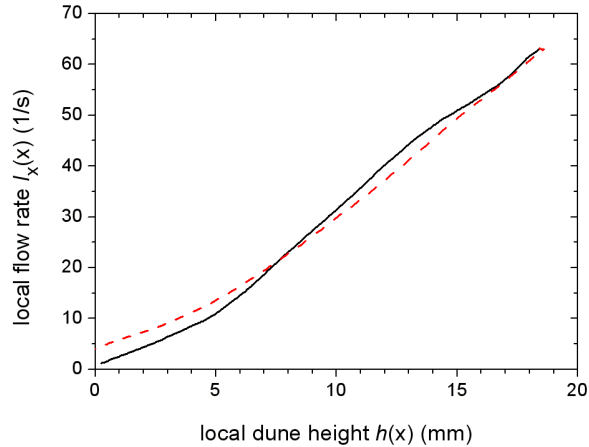}
\caption{\label{fig15}Local relation between the total flow rate $I_{\textnormal{\scriptsize{x,total}}}(x)$ and the height of the dune $h(x)$ for the upstream side (black solid line) and the downstream side (red dashed line).}
\end{center}
\end{figure}

The erosion and deposition of sand leads to a temporal change in the height of the dune surface. In the steady state the dune moves shape invariantly and consequently the relation between changing sand flow rate and the deformation of the dune shape becomes constant. From the conservation of mass a term for the migration velocity of dunes ($v_{\textnormal{\scriptsize{com}}}$ in our experiment) can be derived as \cite{bagnold1941, sauermann2003}
\begin{displaymath}
v_{\textnormal{\scriptsize{com}}} \sim \frac{ \textnormal{d}I_{\textnormal{\scriptsize{x,total}}} }{ \textnormal{d}h } = \textnormal{const.}
\end{displaymath}
In field measurements this simple formula is used to check the present steady state of a desert dune \cite{sauermann2003}. For our model dune the relation of sand flow rate $I_{\textnormal{\scriptsize{x,total}}}(x)$ and dune height $h(x)$ at a certain horizontal position $x$ is plotted in Fig.\,\ref{fig15}. The dune is partitioned at the crest in an upstream side and a downstream side. Above a height of about 4\,mm (two grain diameters) the slope $\textnormal{d}I_{\textnormal{\scriptsize{x,total}}} / \textnormal{d}h$ becomes constant for both sides, an additional confirmation for the steady state.

\begin{figure}
\begin{center}
\includegraphics[width = 0.75\linewidth]{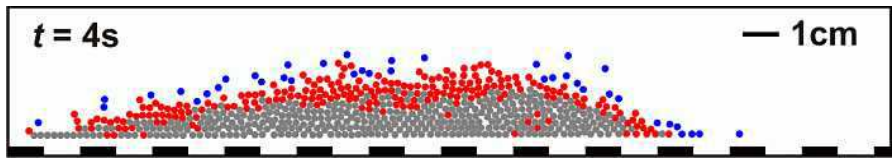}
\caption{\label{fig16}The figure shows a detail of the snapshot at $t = 4$\,s in Fig.\,\ref{fig3}. The detected particles are drawn as disks and color coded with respect to their type: resting particles (grey), creeping particles (red), and saltating particles (blue). See animated movie2 for the corresponding image sequence between $t = 4$\,s and $t = 5$\,s.}
\end{center}
\end{figure}

As mentioned above the bulk of the dune is composed of resting particles and creeping particles. To separate them, we use the threshold velocity $v_{\textnormal{\scriptsize{th}}} = 5$\,mm/s as criteria. The value of $v_{\textnormal{\scriptsize{th}}}$ is given by the standard deviation of the velocity distribution of all beads. Note, $v_{\textnormal{\scriptsize{th}}} \approx 1/10 v_{\textnormal{\scriptsize{com}}}$. It can be seen in Fig.\,\ref{fig16} that sometimes beads in the core of the dune are set in motion, which results from the regrouping of the beads due to the erosion and deposition processes. Moreover, the snapshot in Fig.\,\ref{fig16} illustrates that only a few particles are completely detached from the surface. This corresponds to $\varphi_{\textnormal{\scriptsize{v}}} (x,y)$ shown in Fig.\,\ref{fig8}.

\begin{figure}
\begin{center}
\includegraphics[width = 0.6\linewidth]{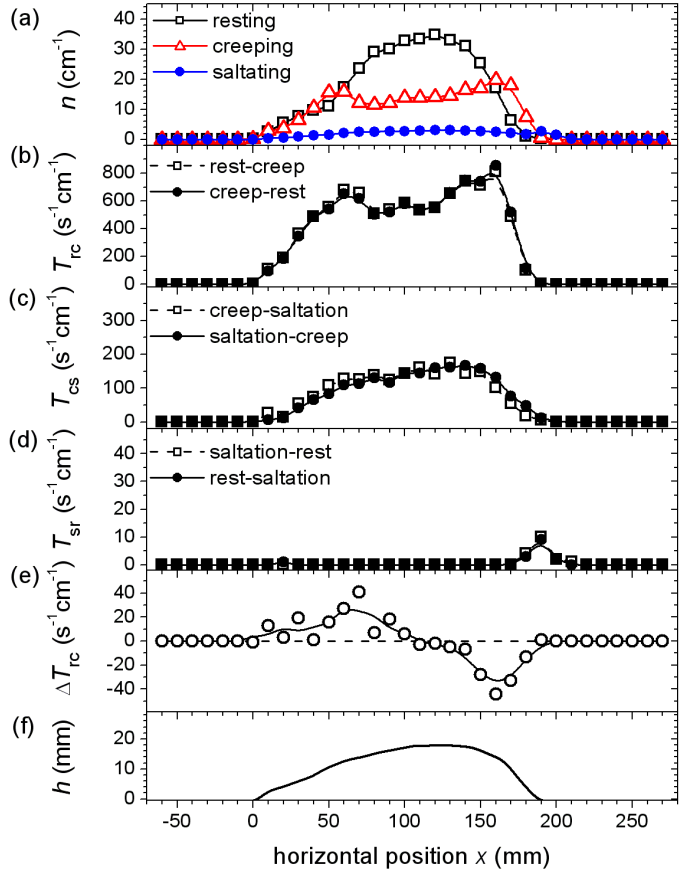}
\caption{\label{fig17} Local particle densities for resting ($n_{\textnormal{\scriptsize{r}}}$), creeping ($n_{\textnormal{\scriptsize{c}}}$), and saltating ($n_{\textnormal{\scriptsize{s}}}$) particles are plotted in (a). Panels (b)-(d) show the pairs of transition rates $T_{\textnormal{\scriptsize{rc}}}$, $T_{\textnormal{\scriptsize{cs}}}$, and $T_{\textnormal{\scriptsize{rs}}}$ between the three types of particles. The difference between the rest-creep and creep-rest transition in (b) gives $\Delta T_{\textnormal{\scriptsize{rc}}}$ in panel (e). The contour of the dune bulk $h$ is shown in panel (f).}
\end{center}
\end{figure}

These observations are quantitatively shown in Fig.\,\ref{fig17}(a) with the distributions of the particle densities $n_{\textnormal{\scriptsize{r}}}$ (resting), $n_{\textnormal{\scriptsize{c}}}$ (creeping), and $n_{\textnormal{\scriptsize{s}}}$ (saltating). For the analysis every recorded image is divided in columns with a width of 1\,cm. The particle densities are determined as the time averaged number of particles counted in each column between $t = 4$\,s and $t = 5$\,s. For technical reasons the frames are shifted according to the center of mass of the dune.

The distributions of $n_{\textnormal{\scriptsize{r}}}$ and $n_{\textnormal{\scriptsize{c}}}$ in Fig.\,\ref{fig17}(a) show an excess of creeping particles on the downstream side, where the deposition occurs. At the upstream bottom of the dune $n_{\textnormal{\scriptsize{c}}}$ and $n_{\textnormal{\scriptsize{s}}}$ increase. They take constant values in the middle section and decrease in the deposition area. Note that the local densities $n_{\textnormal{\scriptsize{c}}}$ and $n_{\textnormal{\scriptsize{s}}}$ show that there are more creeping particles than saltating particles, but Fig.\,\ref{fig14}(b) shows that the flow rate of the saltating beads contribute the larger part of $I_{\textnormal{\scriptsize{x,total}}}(x)$. This can be explained by the higher speed of the saltating particles.

The particle tracking method (see Fig.\,\ref{fig4}) allows for the investigation of transitions between the three types of beads in our experiment. Because every transition has two directions, we have to distinguish between rest-creep, creep-rest, creep-saltation, saltation-creep, rest-saltation, and saltation-rest. The number of transitions per second gives the transition rate. To simplify the notation, we label each connected pair with one variable: $T_{\textnormal{\scriptsize{rc}}}$ (rest and creep), $T_{\textnormal{\scriptsize{cs}}}$ (creep and saltation), and $T_{\textnormal{\scriptsize{rs}}}$ (rest and saltation). The local distributions of the transition rates corresponding to Fig.\,\ref{fig17}(a) are shown in Fig.\,\ref{fig17}(b)-(d). In the areas of erosion and deposition $T_{\textnormal{\scriptsize{rc}}}$ takes a local maximum. The maximum of $T_{\textnormal{\scriptsize{cs}}}$ lies at the crest. The high values of all transition rates indicate the frequent collisions between the beads during the dune migration. The distribution of $T_{\textnormal{\scriptsize{rs}}}$ in Fig.\,\ref{fig17}(d) shows that no reptation in the aeolian sense occurs in our experiment. The saltating beads do not originate from resting beads. They originate from creeping beads dragged from the fluid. The peak at the downstream side results from single beads, which are temporally not connected to the bulk of the dune. The net rate of erosion and deposition is obtained by subtracting the two parts of $T_{\textnormal{\scriptsize{rc}}}$, which yields $\Delta T_{\textnormal{\scriptsize{rc}}}$ in Fig.\,\ref{fig17}(e). The resulting curve fits to the erosion rate shown in Fig.\,\ref{fig14}(c). The distribution of $n_{\textnormal{\scriptsize{r}}}$  in panel (a) reflects the contour of the bulk, shown in Fig.\,\ref{fig17}(f).

To conclude, we designed an experiment which allows the direct measurement of the particle dynamics and the surrounding driving water flow above and inside a two-dimensional barchan dune model. The measurement of the particle dynamics inside the dune reveals the nature of the crawling motion of dunes, whose speed is one order of magnitude smaller than that of the creep and saltation of individual grains. This ratio is of course not universal, but specific for our downsized model dune. Due to the possibility to distinguish between creeping and saltating particles, we show that the erosion rate consists of comparable contributions from both. The saltation flow rate is slightly larger, whereas the number of saltating particles is one order of magnitude lower than that of the creeping ones. The velocity field of the saltating particles is comparable to the velocity field of the driving fluid, although that of the particles lags behind.

Moreover, it is demonstrated that the initial triangular heap evolves to a steady state with constant mass, shape, velocity, and packing fraction after one turnover time has elapsed. Within that time the mean distance between particles initially in contact reaches a value of approximately one quarter of the dune basis length, a number which serves to quantify the mixing process.

The shear stress of the water flow is shown to be related to the particle erosion rate, as well as the slope of the dune surface. It can be observed that the spatial profile of the shear stress reaches its maximum value upstream of the crest, while its minimum lies at the downstream foot of the dune. The particle tracking method reveals that the deposition of entrained particles occurs primarily in the region between these two extrema of the shear stress. Due to this deposition most of the mass of the dune is preserved. The upstream shift of the shear stress has been a crucial element for explaining the steady state in the minimal models. For a more fundamental understanding of the dune dynamics the presented facts are believed to encourage particulate models and simulations.

It's a pleasure to thank Kai Huang and Matthias Schr\"{o}ter for stimulating suggestions. We are grateful for support from Deutsche Forschungsgemeinschaft through Kr1877/3-1 (Forschergruppe 608 'Nichtlineare Dynamik komplexer Kontinua').

\section*{References}

\bibliographystyle{unsrt}

\end{document}